\documentclass[aps,prb,twocolumn,superscriptaddress]{revtex4-2}
\usepackage[colorlinks=true,citecolor=blue,linkcolor=blue,urlcolor=blue]{hyperref}
\usepackage{graphicx, nicefrac, textcomp}
\usepackage[dvipsnames]{xcolor}
\usepackage{hyperref}
\pdfoutput=1

\begin{document}

\title{Topotactically induced oxygen vacancy order in nickelate single crystals} 

\author{Yu-Mi Wu}
\email{yu-mi.wu@fkf.mpg.de}
\author{Pascal Puphal}
\author{Hangoo Lee}
\author{Jürgen Nuss}
\author{Masahiko Isobe}
\author{Bernhard Keimer}
\author{Matthias Hepting}
\affiliation{Max Planck Institute for Solid State Research, Heisenbergstrasse 1, Stuttgart 70569, Germany}
\author{Y. Eren Suyolcu}
\email{eren.suyolcu@fkf.mpg.de}
\affiliation{Max Planck Institute for Solid State Research, Heisenbergstrasse 1, Stuttgart 70569, Germany}
\affiliation{Department of Materials Science and Engineering, Cornell University, Ithaca, New York 14853, USA}
\author{Peter A. van Aken}
\affiliation{Max Planck Institute for Solid State Research, Heisenbergstrasse 1, Stuttgart 70569, Germany}

\begin{abstract}
The strong structure-property coupling in rare-earth nickelates has spurred the realization of new quantum phases in rapid succession. Recently, topotactic transformations have provided a new platform for the controlled creation of oxygen vacancies and, therewith, for the exploitation of such coupling in nickelates. Here, we report the emergence of oxygen vacancy ordering in Pr$_{0.92}$Ca$_{0.08}$NiO$_{2.75}$ single crystals obtained via a topotactic reduction of the perovskite phase Pr$_{0.92}$Ca$_{0.08}$NiO$_{3}$, using CaH$_2$ as the reducing agent. We unveil a brownmillerite-like ordering pattern of the vacancies by high-resolution scanning transmission electron microscopy, with Ni ions in alternating square-pyramidal and octahedral coordination along the pseudocubic [100] direction. 
Furthermore, we find that the crystal structure acquires a high level of internal strain, where wavelike modulations of polyhedral tilts and rotations accommodate the large distortions around the vacancy sites.
Our results suggest that atomic-resolution electron microscopy is a powerful method to locally resolve unconventional crystal structures that result from the topotactic transformation of complex oxide materials.

\end{abstract}
\pacs{df}
\maketitle

\section{Introduction}
In transition metal oxides, strongly correlated valence electrons can couple collectively to the lattice degrees of freedom, which can lead to a variety of emergent ordering phenomena, including exotic magnetism, multiferroicity, orbital order, and superconductivity \cite{khomskii2014transition}. In oxides with the perovskite structure, high flexibility and tolerance to structural and compositional changes enable the controlled exploitation and manipulation of the emergent properties \cite{torrance1992systematic}. Oxygen vacancies, for example, can radically alter the electronic states in materials, and in turn, suppress or enhance emergent phases via charge compensation and/or structural phase transitions \cite{Zimmermann2003,jeong2013,yao2017}. Understanding the formation of oxygen vacancies and their impact thus provides promising prospects for exploring new physical properties and potential future technological applications.

A prototypical example for correlated transition-metal oxides is the family of perovskite rare-earth nickelates, $R$NiO$_3$ ($R$ = rare-earth ion), exhibiting a rich phase diagram including metal-to-insulator and antiferromagnetic transitions \cite{catalan2008progress,Middey2016,catalano2018rare}. For $R$ = Pr and Nd, these transitions occur concomitantly with a breathing distortion of the NiO$_6$ octahedra and a disproportionation of the Ni-O hybridization \cite{Johnston2014,mercy2017structurally,suyolcu2021control,lu2016}. As a consequence, the material family exhibits a pronounced structure-property relationship \cite{Boris2011,Chakhalian2011,Hepting2014,Disa2015,Fabbris2016,Hepting2018,Fowlie2019} and a sensitivity to oxygen vacancy formation, which can modify the surrounding Ni-O bonds and the nominal 3$d^{7}$ electronic configuration of the Ni$^{3+}$ ions \cite{li2022doping}. Notably, an extensive oxygen reduction of the perovskite phase towards Ni$^{1+}$ with a cuprate-like 3$d^{9}$ electronic configuration was recently realized via topochemical methods in Nd$_{0.8}$Sr$_{0.2}$NiO$_2$ thin films, yielding the emergence of superconductivity \cite{li2019superconductivity}. Furthermore, superconductivity was also observed in topotactically reduced films with $R$ = La and Pr \cite{Osada2020,osada2021,Wang2022}, as well as for various Sr-substitution levels \cite{zeng2020phase,Li20201} and substitution with Ca ions \cite{zeng2021}. Since these reduced nickelates with the infinite-layer crystal structure are nominally isoelectronic and isostructural to cuprate superconductors, the degree of the analogy between the two material families is vividly debated \cite{Hepting2021,botana2020similarities,rossi2021orbital,goodge2021doping,gu2020single,been2021electronic}. Moreover, vigorous  efforts are ongoing to realize superconductivity not only in thin films, but also in polycrystalline powders \cite{Li2020powder,Wang2020powder} as well as in single-crystalline samples \cite{puphal2021,Puphal2022b}, while also an improved understanding of the topotactic reduction process between the perovskite and infinite-layer phase is highly desirable. In particular, the reduction involves various intermediate (metastable) phases, in which the oxygen vacancy ordering patterns and the nature of the emergent phases have not yet been clarified comprehensively. 

For instance, extensive experimental and theoretical studies \cite{alonso1997structural,sanchez1996metal,abbate2002electronic,malashevich2015first,misra2016transport,Nguyen2020,shin2022magnetic} were performed on oxygen deficient LaNiO$_{3-\delta}$ with $0 < \delta \leq 0.5$, suggesting a transition from a paramagnetic metal to a ferromagnetic semiconductor and an antiferromagnetic insulator as a function of increasing vacancy concentration \cite{wang2018antiferromagnetic,liao2021oxygen,misra2016oxygen}. 
For $\delta \approx 0.5$, neutron powder diffraction \cite{alonso1997structural} revealed that the parent perovskite crystal structure with uniform NiO$_6$ octahedra changed to a structure with sheets of NiO$_6$ octahedra and square-planar NiO$_4$ units arranged along the pseudocubic [100] direction \cite{alonso1997structural,sanchez1996metal}, involving a 2a$_{p}\times$2a$_{p}\times$2a$_{p}$ reconstruction of the parent pseudocubic unit cell (a$_{p}$ is the pseudocubic lattice parameter). Yet, the detailed crystal structure for the case $\delta \approx 0.25$ is not known, although an electron diffraction study suggested that it involves a  2a$_{p}\sqrt{2}\times$2a$_{p}\sqrt{2}\times$2a$_{p}$ reconstructed supercell \cite{Gonzales1989}. 
Moreover, for compounds with $R$ = Pr and Nd, even less understanding of the oxygen deficient phases exists. Metastable structures with ferromagnetic order were initially identified for $\delta \approx 0.7$, with x-ray diffraction data indicating a  3a$_{p}$$\times$a$_{p}$$\times$3a$_{p}$ supercell that possibly comprises two sheets of NiO$_4$ square-planar units connected with one sheet of NiO$_6$ octahedra \cite{moriga1994reduction}. In a subsequent neutron powder diffraction study it was suggested, however, that the metastable phase of the Pr compound rather corresponds to $\delta \approx 0.33$ with a $\sqrt{5}$a$_{p}\times$a$_{p}\times\sqrt{2}$a$_{p}$ reconstruction and one sheet of NiO$_4$ square-planar units connected with two sheets of NiO$_6$ octahedra \cite{moriga2002reduction}.

Here, we use atomic-resolution scanning transmission electron microscopy (STEM) together with electron energy-loss spectroscopy (EELS) to investigate the oxygen vacancy formation occurring in a Pr$_{0.92}$Ca$_{0.08}$NiO$_{3-\delta}$ single crystal upon topotactic reduction. We resolve the chemical composition and the atomic-scale lattice of the crystal, identifying a  4a$_{p}$$\times$4a$_{p}$$\times$2a$_{p}$ reconstructed superstructure with a highly distorted Pr sublattice. We find that the oxygen vacancy ordering pattern corresponds to a brownmillerite-like structure with a two-layer-repeating stacking sequence of NiO$_6$ octahedra and NiO$_5$ square pyramids, suggesting an oxygen deficiency of $\delta \approx 0.25$. Meanwhile, quantification of the octahedral tilts and Ni-O bond angles reveals distinct periodic wavelike patterns of polyhedra coordination in different layers due to the oxygen vacancies. These results are markedly distinct from previous reports on reduced rare-earth nickelates and provide an atomic-scale understanding of the moderately oxygen deficient structure with $\delta \approx 0.25$, which is one of the metastable phases occurring during the topotactic reduction process towards the infinite-layer phase of the superconducting nickelates with $\delta = 1$.

\section{Methods}

Single crystals of perovskite Pr$_{1-x}$Ca$_{x}$NiO$_{3}$ were synthesized under high pressure and high temperature. Specifically, a 1000 ton press equipped with a Walker module was used to realize a gradient growth under a pressure of 4 GPa, executed in spatial separation of the oxidizing KClO$_4$ and NaCl flux, similarly to the previous synthesis of La$_{1-x}$Ca$_{x}$NiO$_{3}$ single crystals \cite{puphal2021}. The precursor powders were weighed in according to a desired composition of Pr$_{0.8}$Ca$_{0.2}$NiO$_3$, although the incorporated Ca content in the obtained Pr$_{1-x}$Ca$_{x}$NiO$_{3}$ was lower and ranged from $x = 0.08$ to 0.1. The Pr$_{1-x}$Ca$_{x}$NiO$_{3}$ single crystals were reduced  using CaH$_2$ as the reducing agent in spatial separation to the crystals. The duration of the reduction was eight days, using the same procedure and conditions as previously described for the reduction of La$_{1-x}$Ca$_{x}$NiO$_{3}$ crystals \cite{puphal2021}. 

Single-crystal x-ray diffraction (XRD) was performed on crystals before and after the reduction. The technical details are given in the Supplemental Material \cite{supp}. 

Electron-transparent TEM specimens of the sample were prepared on a Thermo Fisher Scientific focused ion beam (FIB) using the standard liftout method. Samples with a size of 20 $\times$ 5 $\mu$m$^2$ were thinned to 30 nm with 2 kV Ga ions, followed by a final polish at 1 kV to reduce effects of surface damage.
HAADF, ABF and EELS were recorded by a probe aberration-corrected JEOL JEM-ARM200F scanning transmission electron microscope equipped with a cold-field emission electron source and a probe Cs corrector (DCOR, CEOS GmbH), and a Gatan K2 direct electron detector was used at 200 kV. STEM imaging and EELS analyses were performed at probe semiconvergence angles of 20 and 28 mrad, resulting in probe sizes of 0.8 and 1.0 Å, respectively. Collection angles for STEM-HAADF and ABF images were 75 to 310 and 11 to 23 mrad, respectively. To improve the signal-to-noise ratio of the STEM-HAADF and ABF data while minimizing sample damage, a high-speed time series was recorded (2 $\mu$s per pixel) and was then aligned and summed. 
STEM-HAADF and ABF multislice image simulations of the crystal along [100] and [101] zone axis were performed using the QSTEM software \cite{koch2002determination}. Further details of the parameters used  for the simulations are given in the Supplemental Material \cite{supp}.

\section{Results}

\begin{figure*}[t]
\centering
\includegraphics[width=\textwidth]{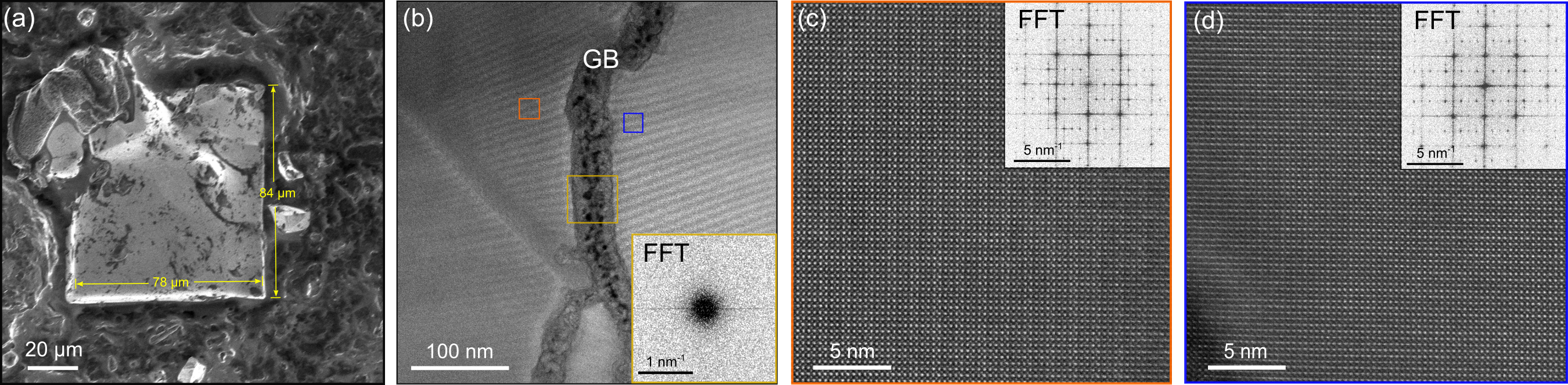}
\caption{(a) SEM-secondary electron (SE) image of a Pr$_{0.92}$Ca$_{0.08}$NiO$_{2.75}$ crystal. (b) STEM-HAADF image taken at low magnification showing that grain boundary (GB) like regions separate single-crystalline regions. Inset shows the fast Fourier transform pattern from the yellow square on the GB. (c),(d) Atomic-resolution images of the regions from the orange and blue squares in (b), respectively. Insets are corresponding fast Fourier transform patterns, indicating that the same structural phase is present in both regions.}
\end{figure*}

In thin films of infinite-layer nickelates, the highest superconducting transition temperatures are realized through a substitution of approximately 20\% of the rare-earth ions by divalent Sr or Ca ions \cite{zeng2020phase,Li20201,zeng2021}. Accordingly, we have prepared the precursor materials for the synthesis of single crystals with a nominal stoichiometry of Pr$_{0.8}$Ca$_{0.2}$NiO$_3$. Using a high-pressure synthesis methods \cite{puphal2021}, we obtain Pr$_{1-x}$Ca$_{x}$NiO$_{3}$ crystals with typical lateral dimensions of 20 -- 100 $\mu$m. Yet, an analysis of the as-grown crystals by scanning electron microscopy (SEM) with energy-dispersive x-ray (EDX) spectroscopy indicates that the incorporated Ca content lies between $x = 0.08$ and 0.1 (see Fig. S1 in the Supplemental Material \cite{supp}). This discrepancy to the nominal Ca content suggests that different growth parameters, such as an increased oxygen partial pressure, might be required to achieve stoichiometric Pr$_{0.8}$Ca$_{0.2}$NiO$_3$ crystals. By contrast, the employed parameters yielded higher Ca substitutions as high as $x = 0.16$ in the case of La$_{1-x}$Ca$_{x}$NiO$_{3}$ as determined on the as-grown crystal surface by EDX 
\cite{puphal2021}, which exhibits a less distorted perovskite structure \cite{catalan2008progress}. 

As a next step, we use single-crystal XRD to investigate a 20 $\mu$m piece that was broken off from a larger as-grown crystal. The acquired XRD data indicate a high crystalline quality (see Fig. S2 of the Supplemental Material \cite{supp}) and can be refined in the orthorhombic space group $Pbnm$, which is consistent with PrNiO$_3$ single crystals and polycrystalline powders \cite{garcia1992neutron,zheng2019high}. The refined Ca content of the crystal is 8.6\%. The refined lattice parameters and atomic coordinates are presented in the Supplemental Material \cite{supp}. Furthermore, we find that the investigated crystal piece contains three orthorhombic twin domains extracted from the refinement, with volume fractions 0.95/0.04/0.01.

Subsequently, we carry out the topotactic oxygen reduction on a batch of Pr$_{1-x}$Ca$_{x}$NiO$_{3}$ single crystals for eight days, using the same conditions as previously described for the reduction of La$_{1-x}$Ca$_{x}$NiO$_{3}$ crystals \cite{puphal2021}. Single-crystal XRD measurements on a reduced 20 $\mu$m crystal indicate a significant transformation of the crystal structure after eight days. However, a strong broadening and the resulting overlap of the Bragg reflections in the XRD maps prohibit a structural refinement and determination of the symmetry by this method (see Fig. S2 \cite{supp}).

Hence, in order to investigate the topotactic transformation of the crystal lattice on a local scale, we turn to atomic-resolution STEM imaging. We examine a reduced Pr$_{0.92}$Ca$_{0.08}$NiO$_{3-\delta}$ crystal with lateral dimensions of $\sim$80 $\mu$m. A top-down view of the crystal is shown in Fig. 1(a). Identical TEM specimens were prepared from a region of the crystal without visible surface defects caused by the FIB process.
Figure 1(b) shows a low-magnification STEM high-angle annular dark-field (HAADF) image. As a first characteristic of the topotactic-reduced crystal, we note that single-crystalline regions in the specimen are separated by grain boundary (GB) like regions,  with a width ranging from a few ten to hundred nanometers  and a length ranging from a few nanometers to micrometers. The GBs exhibit mostly an amorphous structure [Fig. 1(b) inset and Fig. S3], exhibiting dark contrast in the images that originate from diffuse scattering \cite{WANG1994} (see Fig. S3 for more details \cite{supp}). The amorphous character of the GBs is also confirmed in EELS measurements of elemental distribution profiles across a GB, which show a reduced EELS intensity of all cations due to the deteriorating signal in structurally disordered regions (Fig. S4 \cite{supp}). The presence of GBs can be a consequence of topotactic reduction. Alternatively, the GBs might have formed already during the high-pressure growth of the perovskite phase.

Zoom-in STEM-HAADF images from areas on either side of a GB [orange and blue squares in Fig. 1(b)] are displayed in Figs. 1(c) and 1(d). The same lattice structure and orientation are observed from the different crystalline domains near the GB.
One typical domain size in the crystal is found to be hundreds of nanometers. STEM-HAADF imaging over larger field of view of hundred nanometers does not reveal any regions with defects or impurity phase inside one crystalline domain, suggesting that each domain retains a high crystalline quality and a stable structural phase after the reduction process (see Fig. S3 \cite{supp}).

\begin{figure*}[t]
\centering
\includegraphics[width=5.0625in]{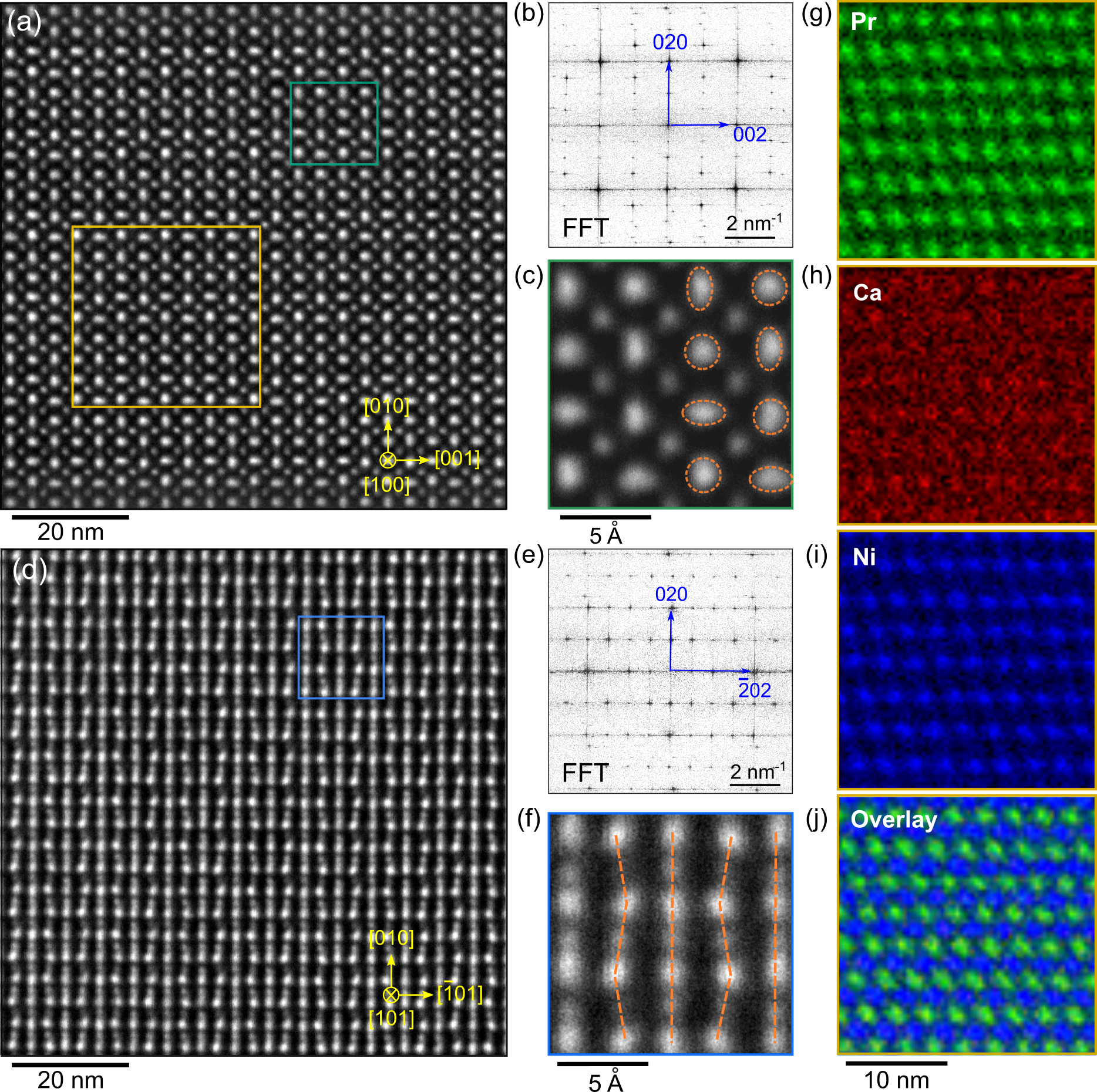}
\caption{(a),(d) STEM-HAADF images viewed along the pseudocubic [100] and [101] directions, respectively. (b),(e) The fast Fourier transformation (FFT) of the HAADF images along two projection directions. The presence of superstructure reflexes indicates the formation of a fourfold superstructure in real space. (c),(f) Zoom-in views of the images from the green and blue rectangles in (a) and (d), respectively. The projection of the column of Ni atoms exhibits a round shape, whereas the Pr atoms appear elongated along the [100] direction and zigzag along the [101] direction. (g)-(j) STEM-EELS elemental maps along the [100] direction of Pr, Ca, Ni, and the overlaid map, obtained from the yellow square in (a).}
\end{figure*}

Figure 2(a) and 2(d) show atomic-resolution STEM-HAADF images viewed along the [100] and [101] pseudocubic orientations, respectively.
The corresponding fast Fourier transforms (FFT) of the images reveal the crystal symmetry [Figs. 2(b) and 2(e)]. As presented in Fig. 2(b), there is an approximate 8\% difference of the distances between the FFT maxima in reciprocal space along the [002] and [020] axis (blue arrows on FFT patterns). Based on the preference of vacancy formation on the apical oxygen sites \cite{alonso1995preparation}, this indicates a contraction along the [002] axis due to the removal of apical oxygen atoms following the topotactic reduction process. Between the FFT maxima, the satellite spots corresponding to the superstructure periodicity appear. The wave vector of the superstructure reflexes is $q$ = 1/4 reciprocal lattice units along [020] axis and $q$ = 1/2 along [002] axis, indicating a formation of the 4a$_{p}$$\times$4a$_{p}$$\times$2a$_{p}$ superstructure of perovskite [$a$ = $b$ = 16.56(32) \AA, $c$ = 7.60(14) \AA].
In STEM-HAADF images, the contrast is approximately proportional to $Z^{1.7-2}$ (where $Z$ is the atomic number) \cite{treacy1978z,pennycook1988chemically}, so Pr columns ($Z$ = 59) appear bright and Ni columns ($Z$ = 28) exhibit a darker contrast.
The fourfold superstructure arises from the periodic changes in the intensity of the STEM image. As shown in the zoomed-in image [Fig. 2(c)], half of the Pr atoms along the [100] projection appear elongated, while the other half remain round and undistorted. Focusing on the first two Pr rows, the atoms exhibit alternating round and vertical oval shapes, while the third and fourth rows exhibit alternating round and horizontal oval shapes. 

Considering the projection in our image, the distortions stem from displacements of Pr columns. Figure 2(d) indeed reveals straight and zigzag line patterns of Pr atoms alternating along the [$\bar{1}$01] direction. A closer look at the atomic arrangement in Fig. 2(f) shows that the zigzag line on the fourth column appears to be a mirror of the second, forming a four-layer repeat sequence (ABAC stacking). 
EELS elemental maps of Pr, Ca and Ni obtained from the crystal are displayed in Figs. 2(g)-2-(i) using Pr $M_{5,4}$, Ca $L_{3,2}$, and Ni $L_{3,2}$ edges, respectively. The maps show that the Pr, Ca and Ni contents are homogeneous over the structure. Integrated concentration profiles of Pr and Ca confirm the A-site cation stoichiometry and uniform distribution (see Fig. S5 \cite{supp}). This suggests that strong distortions of the Pr lattice are likely not rooted in an A/B-site deficiency or ordering, but rather originate from other factors such as oxygen vacancy formation. 

\begin{figure*}[t]
\centering
\includegraphics[width=\textwidth]{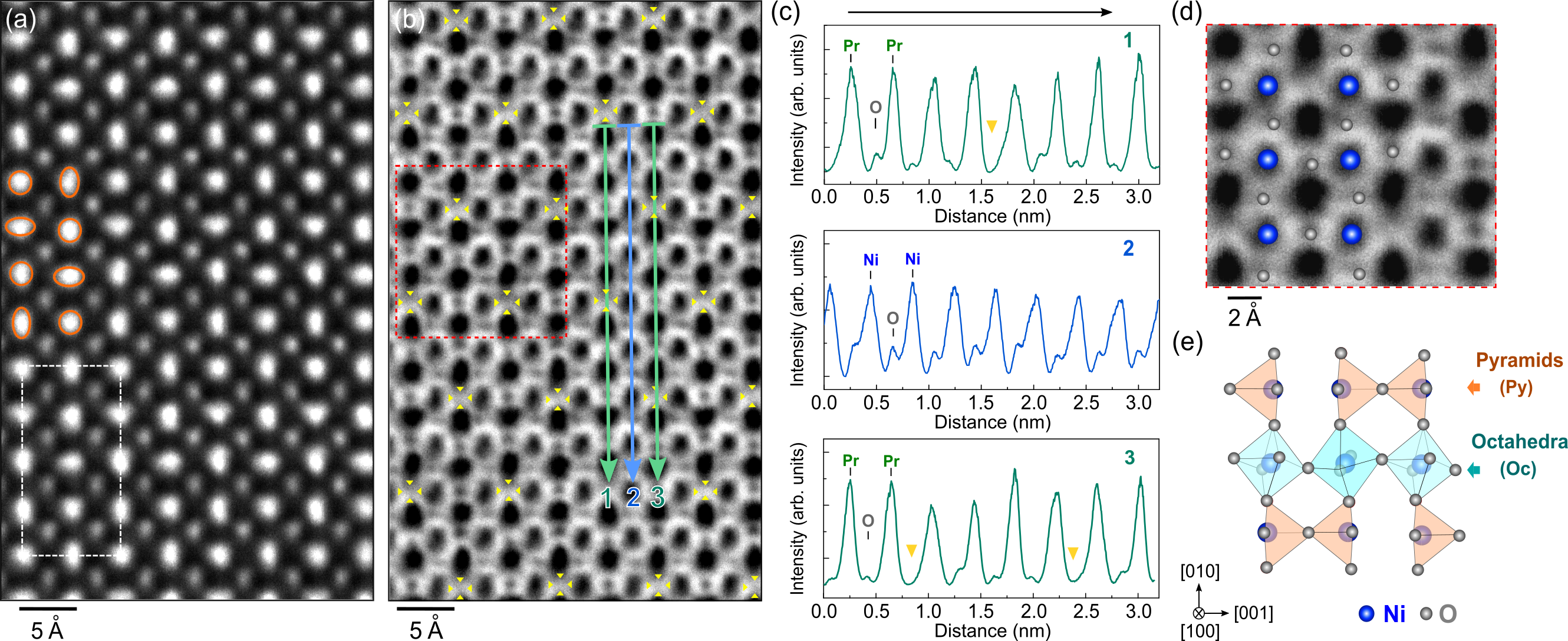}
\caption{(a),(b) The simultaneously acquired STEM-HAADF and ABF images viewed along the [100] direction, respectively. The white dashed rectangle in (a) indicates one reconstructed  supercell. The yellow triangles in (b) indicate the positions of oxygen vacancies. (c) ABF intensity profiles (with inverse intensities) along the green and blue lines indicated in (b), showing the alternating Pr-O and Ni-O sites. Yellow triangles indicate the positions of oxygen vacancies. (d) Zoom-in image of the region indicated by the dashed red rectangle in (b). (e) Sketch of the structural model for the Ni and O sublattice viewed along the [100] direction, with alternating layers of apex-linked NiO$_5$ pyramids in orange and NiO$_6$ octahedra in cyan.}
\end{figure*}

To gain insights into the detailed atomic structure, high-resolution STEM annular bright-field (ABF) images were acquired for imaging lighter elements such as oxygen. Figures 3(a) and 3(b) show the simultaneously acquired HAADF and ABF images of the crystal along the [100] projection. 
The distribution of oxygen ions including filled and empty apical oxygen sites is clearly visible by ABF imaging.
The corresponding inverse intensity profiles extracted from different layers are displayed in Fig. 3(c).
The absence of image contrast at every fourth oxygen site of the Pr-O layers confirms the vacancy ordering (profiles 1 and 3), while the oxygen content remains constant in Ni-O layers (profile 2).
By overlaying the yellow arrows, the ordering pattern of oxygen vacancies can be clearly visualized. Half of the NiO$_6$ octahedra lose one apical oxygen atom and the remaining five oxygen atoms in a pseudocubic unit cell form a NiO$_5$ pyramid.

Notably, a square pyramidal coordination of the Ni ion is uncommon in nickel-oxide based materials. Few compounds with Ni\textsuperscript{2+} ions in similar five-fold coordination include KNi$_4$(PO$_4$)$_3$ and BaYb$_2$NiO$_5$ \cite{waroquiers2017statistical}. However, none of these compounds exhibits a perovskite-derived structure. Our investigation therefore reveals the existence of a new structural motif in marked distinction to the NiO$_4$ square-planar coordination in previous reports on oxygen-deficient perovskite nickelates  \cite{manthiram1999factors,shin2022magnetic}.

A magnified view of the ABF image is shown in Fig. 3(d). The apex-linked pyramids as "bow-tie" dimer units form a one-dimensional chain running along the [001] direction. Such configuration is consistent with the brownmillerite type structure A$_n$B$_n$O$_{3n-1}$ with $n$ = 4 corresponding to the  A$_4$B$_4$O$_{11}$ phase: Layers of apex-linked pyramids are stacked in the sequence ...-\textit{Oc-Py-Oc-Py}-..., where \textit{Oc} denotes a layer containing only octahedra (cyan), and \textit{Py} a layer only pyramids (orange). The ...-\textit{Oc-Py-Oc-Py}-... sequence runs parallel to the [010] axis, so that the \textit{Py} layers are at $1/4$ (001) and $3/4$ (001) planes with a stacking vector $1/2$ [001].
In the pyramidal layers, the remaining oxygen atoms are located at the center of apical sites without causing any tilt of square pyramids. In contrast, the apical oxygen atoms in octahedral layers tend to shift towards the elongated Pr atoms, leading to large octahedral tilts.
A corresponding STEM-ABF image simulation was performed based on the predicted atomic model shown in Fig. 3(e) using the multislice method \cite{koch2002determination} (see Fig. S6 \cite{supp}). 
The simulated image reproduces the vacant sites and distortions well as observed from the STEM measurements, confirming the main crystal structure and alternating-stacking configuration.

\begin{figure*}[t]
\centering
\includegraphics[width=\textwidth]{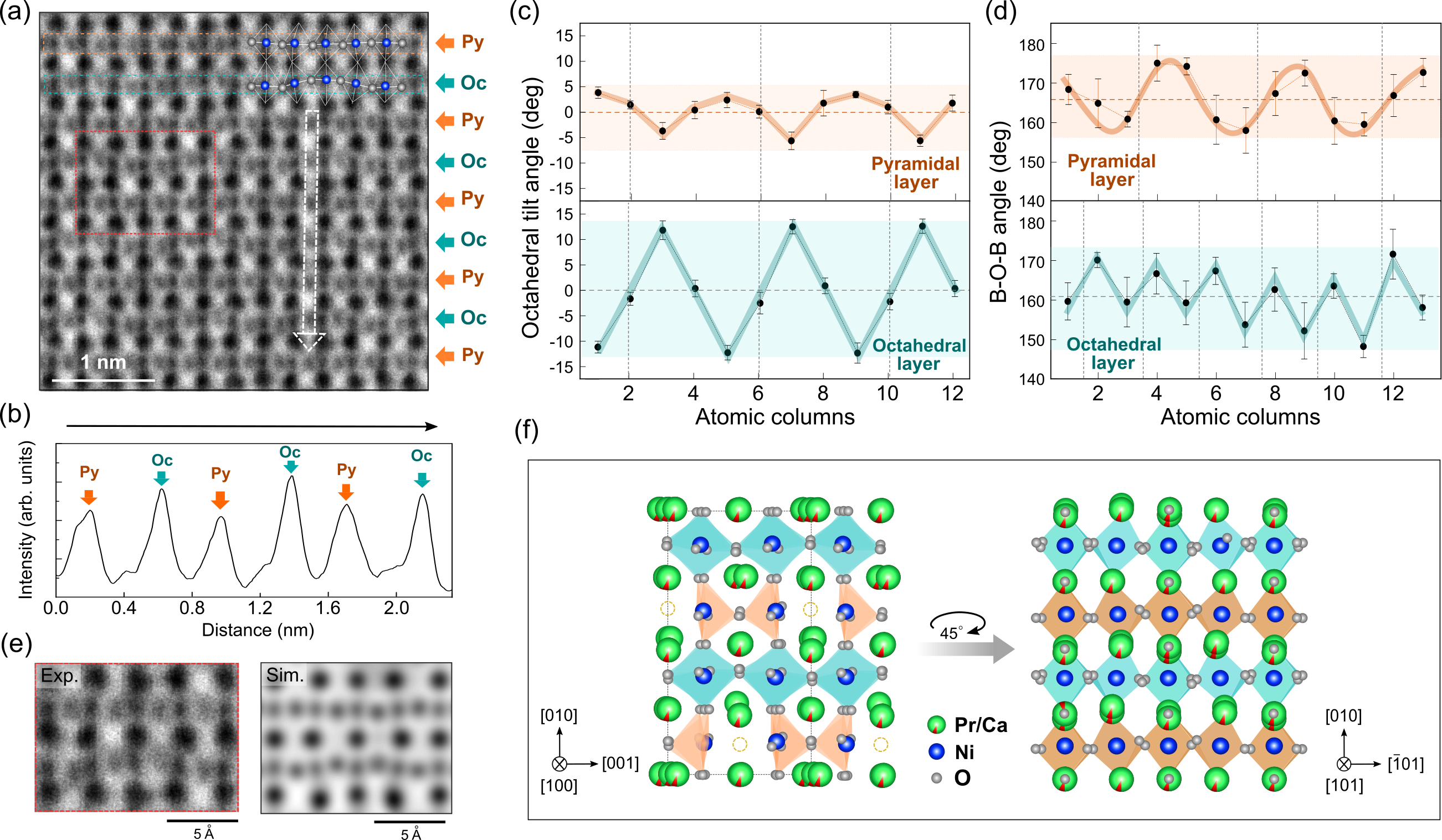}
\caption{(a) A STEM-ABF image viewed along the [101] direction with the overlaid structural model. Ni and O are represented by blue and red dots, respectively. (b) Inverse ABF intensity profile taken from the white dashed rectangle in (a). The higher oxygen intensities correspond to octahedral layers (denoted $Oc$), and lower intensities correspond to pyramidal layers (denoted $Py$). (c),(d) Oxygen octahedral tilt and Ni-O-Ni (B-O-B) bond angle measurements along the pyramidal (P) and octahedral (O) layers, as indicated in (a). The error bars indicate the standard deviation of the mean obtained from 10 rows from the image. The grey dashed lines indicate the averaged values. (e) Left: Zoomed-in image from the red rectangle in (a). Right: Simulated STEM-ABF image along the [101]. (f) Sketch of the structural model of Pr$_{0.92}$Ca$_{0.08}$NiO$_{2.75}$,  viewed along the [100] and [101] directions, respectively. The empty dashed circles indicate the positions of the oxygen vacancies. The black dashed line marks the reconstructed 4a$_{p}$$\times$4a$_{p}$$\times$2a$_{p}$ supercell.}
\end{figure*}

The distribution of oxygen vacancies can lead to modifications in bond angles and tilting of octahedra. Quantitative STEM ABF measurements were used to precisely examine the atomic structure including the oxygen positions along the [101] direction. 
From the inverse ABF intensity profiles taken in Fig. 4(a), as shown in Fig. 4(b), the oxygen intensities on different Ni-O layers vary as a function of the atomic columns along the [010] direction, caused by different apical oxygen occupancies in octahedral and pyramidal layers. Thus, Ni-O layers are denoted \textit{Oc} as the oxygen intensity reaches its maximum on the octahedral layer and denoted \textit{Py} with the lower intensity due to oxygen deficiency on the pyramidal layers.
Analyses of the ABF image, quantifying the NiO$_6$ octahedral tilt angles and Ni-O-Ni bond angles, are shown in Figs. 4(c) and 4(d).
Notably, we observe two different patterns of octahedral distortions along the [$\bar{1}$01] direction on the pyramidal and octahedral layers.
The pyramidal layer shows a variation of tilt angles, with repeating order -4$^\circ$--0$^\circ$--4$^{\circ}$--0$^\circ$ [Fig. 4(c), upper panel]. Meanwhile, a single wavelike pattern of Ni-O-Ni angles varies from 157$^\circ$ to 176$^\circ$ with an average of 166.3$^\circ$ [Fig. 4(d), upper panel]. Such four perovskite subcell wavelike pattern, as indicated by the dashed lines, can arise from the ordering of oxygen vacancies along the [100] projection on the pyramidal layer. 
On the octahedral layer, different from the pyramidal layer, the amplitude of the octahedral tilting is larger, varying from -12$^\circ$ to 12$^\circ$. Ni-O-Ni bond angles on the octahedral layer exhibit a zigzag modulated pattern with two sub-cell repeats and an average of 161.1$^\circ$.
The overall periodicity of bond angles and tilt modulations is consistent with the alternating zigzag and straight pattern on the Pr columns along the [$\bar{1}$01] direction. The simulated ABF image for the predicted model along this viewing direction also agrees well with the STEM-ABF image, confirming the polyhedral distortions in the structure [Fig. 4(e)].
The different amplitudes of tilt and bond angles between pyramidal and octahedral layers are the result of the change in oxygen content. This is also revealed in the structure along the [100] projection [Fig. 3(e)], where NiO$_6$ octahedra show highly distorted tilts in the octahedral layer, while less distorted NiO$_5$ pyramids are present in the pyramidal layer for the lattice accommodation due to the removal of oxygen during reduction.

\section{Discussion and Conclusion}
The obtained insights into the vacancy order in the oxygen sublattice and the distortions of the Pr sublattice are compiled in the two schematics in Fig. 4(f), depicting a brownmillerite-like lattice along the [100] and [101]  investigated in this study. 
Oxygen vacancies at every fourth apical site in every second row lead to a contraction of the out-of-plane lattice parameter and an ordering pattern with alternating NiO$_6$ octahedral and NiO$_5$ pyramidal layers.
We note that the STEM-ABF images indicate a nominal structure of Pr$_{0.92}$Ca$_{0.08}$NiO$_{2.75}$ based on the oxygen vacancy ordering in one crystalline domain, and a possible nonuniform oxygen content variation can occur due to the presence of GBs. This can be the subject of future work to explore the mechanism and origin of GBs \cite{huang2021}.
Compared to the perovskite structure, the vacancy order reduces the tilt angles and increases the Ni-O bond angles in the pyramidal environment. Also the Pr sublattice is affected by the lack of every fourth apical oxygen ion, and the resulting complex distortion pattern leads to a 4a$_{p}$$\times$4a$_{p}$$\times$2a$_{p}$ reconstructed superstructure. 
The observed distortion of the Pr cation position and the associated wavelike variation of the surrounding bond angles and polyhedral tilts is highly unusual for perovskite-related materials. Yet, highly distorted \textit{A}- and \textit{B}-site cation sublattices were also reported for other topotactically reduced perovskite-related materials, such as CaCoO$_2$ \cite{kim2023geometric}, enabling the realization of phases that might be categorically unattainable by direct synthesis methods.

Further studies on Pr$_{0.92}$Ca$_{0.08}$NiO$_{2.75}$ are highly desirable to accurately determine the reconstructed atomic positions and the crystallographic unit cell, which is likely larger than the  4a$_{p}$$\times$4a$_{p}$$\times$2a$_{p}$  supercell. For instance, high-resolution synchrotron XRD might allow to resolve the overlapping structural reflexes in our single-crystal XRD maps (Fig. S2 of the Supplemental Material \cite{supp}) and therewith a full structural refinement might be achievable. Moreover, future STEM studies on crystals after prolonged topotactic reduction can reveal whether our alternating pyramidal/octahedral structure eventually transforms into a square-planar/octahedral structure with a $\sqrt{5}$a$_{p}\times$a$_{p}\times\sqrt{2}$a$_{p}$ supercell, which was proposed in Ref.~\onlinecite{moriga2002reduction} for PrNiO$_{3-\delta}$ with $\delta \approx 0.33$.

Notably, our observed oxygen vacancy ordering with apex-linked pyramidal units is distinct from all previously identified $R$NiO$_{3-\delta}$ lattice structures, which contain stacks of alternating square-planar NiO$_4$ and octahedral NiO$_6$ sheets. Moreover, to the best of our knowledge, pyramidal coordination was generally not observed in perovskite-derived Ni compounds to date, whereas it is a common lattice motif in various oxygen-deficient transition metal oxides, including Fe \cite{TAKEDA1986237,TAKANO1988140}, Co \cite{vidyasagar1984}, and Mn \cite{POEPPELMEIER198271,POEPPELMEIER198289} compounds. In particular, SrFeO$_{3-\delta}$ hosts a variety of oxygen vacancy ordered phases with distinct spin and charge ordered ground states \cite{vidya2006spin,reehuis2012neutron}. Since a closely similar vacancy ordering pattern as in Fig. 4(e) emerges in SrFeO$_{3-\delta}$ for $\delta = 0.25$ (Sr$_4$Fe$_4$O$_{11}$), an exploration of potentially emerging magnetic order in Pr$_x$Ca$_{1-x}$NiO$_{2.75}$ will be of high interest.  

In summary, we examined the topotactic transformation of a Pr$_{0.92}$Ca$_{0.08}$NiO$_{3}$ single crystal to the oxygen-vacancy ordered Pr$_{0.92}$Ca$_{0.08}$NiO$_{2.75}$ phase. The transformed crystal structure contains a 4a$_{p}$$\times$4a$_{p}$$\times$2a$_{p}$ supercell and periodic distortions as well as a zigzag pattern of Pr ions along the [100] and [101] directions, respectively.
The ordering of the oxygen vacancies on the apical oxygen sites forms one-dimensional chains of bow-tie dimer units of NiO$_5$ square pyramids.
These square pyramidal chains run in parallel to the [001] direction, connecting with flattened NiO$_6$ octahedra.
Our atomic-scale observation of the systematic lattice distortions and oxygen vacancies underpins an unexpected pyramidal-type brownmillerite-like phase in the nickelates after a topotactic reduction. Our results are instructive for future efforts to gain a comprehensive understanding of the topotactic reduction of rare-earth nickelates and related materials.

\begin{acknowledgments}
We thank W. Sigle for fruitful discussions and acknowledge T. Heil for valuable technical support and J. Deuschle for TEM specimen preparation. For the crystal growth, the use of the facilities of the Quantum Materials Department of H. Takagi is gratefully acknowledged. This project has received funding in part from the European Union’s Horizon 2020 research and innovation program under Grant Agreement No. 823717–ESTEEM3.
\end{acknowledgments}


%

\end{document}